# Antiferromagnetism of Double Molybdate LiFe(MoO$_4$)$_2$

Meifeng Liu,[1] Yang Zhang,[2] Tao Zou,[3*] V. Ovidiu Garlea,[3] T. Charlton,[3] Yu Wang,[1] Fei Liu,[1] Yunlong Xie,[1] Xiang Li,[1] Lun Yang,[1] Biwen Li,[1] Xiuzhang Wang,[1] Shuai Dong,[2†] Jun-Ming Liu[1,4]

[1]*Institute for Advanced Materials, Hubei Normal University, Huangshi 435002, China*
[2]*School of Physics, Southeast University, Nanjing 211189, China*
[3]*Neutron Scattering Division, Oak Ridge National Laboratory, Oak Ridge, Tennessee 37831, USA*
[4]*Laboratory of Solid State Microstructures and Innovative Center of Advanced Microstructures, Nanjing University, Nanjing 210093, China*

The magnetic properties of the spin-5/2 double molybdate LiFe(MoO$_4$)$_2$ have been characterized by heat capacity, magnetic susceptibility, and neutron powder diffraction techniques. Unlike the multiferroic system LiFe(WO$_4$)$_2$ which exhibits two successive magnetic transitions, LiFe(MoO$_4$)$_2$ undergoes only one antiferromagnetic transition at $T_N$ ~ 23.8 K. Its antiferromagnetic magnetic structure with the commensurate propagation vector $\boldsymbol{k}$ = (0, 0.5, 0) has been determined. Density functional theory calculations confirm the antiferromagnetic ground state and provide a numerical estimate of the relevant exchange coupling constants.

[*] Email: taozoucn@gmail.com

[†] Email: sdong@seu.edu.cn

## I. Introduction

Double tungstates and molybdates with the chemical formula $A^{I}B^{III}(MO_4)_2$ form a big family of transition metal oxides, where $A$ is an alkali metal; $B$ is a trivalent cation such as $Bi^{3+}$, $In^{3+}$, $Fe^{3+}$ or a rare earth $Sm^{3+}$, $Eu^{3+}$; and $M$ represents $W^{6+}$ or $Mo^{6+}$ [1,5]. The combinations of three categories of metal ions make this family full of varieties in physics and functionality. The inclusion of Li ion at $A$ site make them suitable to be used in lithium-ion batteries as anode materials, while the W/Mo can provide a relative large spin-orbit coupling considering its $4d/5d$ orbitals [1, 6-9]. Most importantly, the $B$-site ions with narrow bands can provide other applicable physical properties, such as magnetism and luminescence [10-15].

Recently, $NaFe(WO_4)_2$ was reported to exhibit an incommensurate spiral spin order at low temperature (<4 K) [16], although this magnetism can not induce a net ferroelectric (FE) polarization ($P$) due to the opposite chirality coexisting in this material. Differently, Liu *et al.* revealed a more interesting magnetic spiral in $LiFe(WO_4)_2$, which breaks the spatial reversal symmetry and induces a net FE $P$ along the [010] axis below 19.7 K through the inverse Dzyaloshinskii-Moriya (DM) interaction [17]. Thus $LiFe(WO_4)_2$ is the second experimentally confirmed multiferroic material in the tungstate family, following the first one $MnWO_4$ [18, 19]. Despite the common chemical formula, the crystalline structures of double tungstates/molybdates can vary in a large range. In fact, $LiFe(WO_4)_2$ and $NaFe(WO_4)_2$ are different in their space groups ($C2/c$ vs $P2/m$) and the arrangements of Fe ions are distinct. Such structural diversity makes it is possible to find more exotic magnetic properties in double tungstates/molybdates. For instance, it was reported that $RbFe(MoO_4)_2$ owns a noncollinear magnetic order below 3.8 K which can trigger the ferroelectricity [20]. In addition, there are lots of other tungstates/molybdates members, e.g. $FeWO_4$, $CoWO_4$, $NiWO_4$, $CuWO_4$, $\alpha$-$FeMoO_4$, $\alpha$-$CoMoO_4$ and $NaCr(WO_4)_2$, all of which display collinear antiferromagnetic orders and thus are not ferroelectric [21-25]. Recently, Chen *et al.* synthesized $LiFe(MoO_4)_2$ and investigated for its applications in lithium-ion batteries [6]. Distinct from either $LiFe(WO_4)_2$ or $NaFe(WO_4)_2$, $LiFe(MoO_4)_2$ possesses a new space group $P$-1 and another type of Fe framework. Its excellent electrochemical properties have been carefully studied. Nevertheless, its magnetic properties have never been studied yet.

In this work, we will report the magnetism of LiFe(MoO$_4$)$_2$ determined from combined studies of heat capacity, magnetic susceptibility, neutron powder diffraction, as well as density functional theory (DFT) calculation. An antiferromagnetic (AFM) phase transition is found around 23.8 K. Below the Néel temperature ($T_N$), a long-range magnetic ordering is established with a commensurate propagation vector (0, 0.5, 0), which does not yield multiferroicity. Above $T_N$, short range magnetic correlation persists, leading to considerable magnetic entropy.

**II. Methods**

High quality polycrystalline LiFe(MoO$_4$)$_2$ samples were synthesized using the conventional solid state reaction method in air, with the highly purified powder of oxides and carbonates as starting materials. The stoichiometric mixtures were grounded and fired at 550 °C for 24 hours in air. The resultant powder was re-grounded and pelletized under a pressure of 1000 psi into disks of 2.0 cm in diameter, and then these pellets were sintered at 650 °C for 24 hours in air again. Phase purity of the sample was checked using X-ray diffraction (XRD) with the Cu *K*α radiation at room temperature. The magnetic properties were measured using Superconducting Quantum Interference Device Magnetometer (SQUID) equipped on Quantum Design Magnetometer (MPMSXL-7). The specific heat was measured on the Physical Property Measurement System (PPMS, Quantum Design) using the heat relaxation method. Neutron powder diffraction (NPD) patterns were collected with neutron wavelength *λ* = 2.41 Å at the HB-2A powder diffractometer at High Flux Isotope Reactor, Oak Ridge National Laboratory.

DFT calculations were performed using the Vienna *ab initio* simulation package (VASP) with the projector augmented-wave (PAW) potentials [26-28]. The Perdew-Burke-Ernzerhof for solids (PBEsol) exchange function has been adopted [29] to obtain an accurate description of the crystal structure of LiFe(MoO$_4$)$_2$. The Hubbard $U_{eff}$ (=$U$-$J$) was imposed on Fe's *d* orbitals using the Dudarev implementation [30] considering the strong correlation effect of 3$d$ orbitals. No $U_{eff}$ is applied to Mo's 4$d$ orbitals considering its empty occupation and weaker correlation comparing with 3$d$ orbitals. The cutoff energy of the plane wave basis was fixed to 650 eV due to a quite high convergent value of Li. The Brillouin zone was adopted using 6×6×3 Monkhorst-Pack $k$-point mesh for the minimal magnetic unit cell. Besides, both the lattice constants and atomic positions were fully relaxed until the Hellmann-Feynman force on each atom is below 0.01 eV/Å.

### III. Results and Discussion

Figures 1(a) shows the crystal structure of LiFe(MoO$_4$)$_2$, which is described in the triclinic space group $P$-1 (No. 2) and consisted of separated layers of [LiO$_6$] monocapped trigonal bipyramids, [FeO$_6$] octahedra, and [MoO$_4$] tetrahedra. The framework of magnetic Fe ions in LiFe(MoO$_4$)$_2$ is shown in Fig. 1(b).

Figure 1(c) shows powder XRD pattern of LiFe(MoO$_4$)$_2$ at room temperature. The lattice parameters were refined using the space group $P$-1 with the Rietveld refinement technique. No impurity phase is observed in the XRD power pattern. The refined lattice parameters of LiFe(MoO$_4$)$_2$ are $a$=6.7766 Å, $b$=7.1679 Å, $c$=7.3104 Å, $\alpha$=90.89°, $\beta$=110.38°, $\gamma$ = 105.21°, which are in good consistent with previous works [31].

Besides XRD, NPD data of LiFe(MoO$_4$)$_2$ were collected at $T$=40 K, as shown in Fig. 1(d). The nuclear structure refinement was conducted using FullProf with the Rietveld method[26]. Our refinement confirms that the compound possesses a triclinic structure with the space group $P$-1. No impurity phase was detected. The corresponding lattice parameters are $a$ = 6.751 (1) Å, $b$ = 7.203 (2) Å, $c$ = 7.1702(1) Å, $\alpha$ = 90.6739 (17)°, $\beta$ = 110.2274(14)°, $\gamma$ = 105.5793(17)°, in agreement with above XRD results. Table I summarizes more parameters such as atomic coordinates and displacement parameters.

Figure 2(a) depicts the magnetic susceptibility ($\chi$) and its inverse (1/$\chi$) as a function of temperature ($T$) measured following zero field cooling (ZFC) and field cooling (FC) processes at 0.1 Tesla field. The two curves almost overlap and no bifurcation is seen in the whole temperature range, suggesting the absence of glass behavior. A peak of $d\chi/dT$ was observed around $T_N$~23.8 K, indicating the establishment of long range magnetic ordering, as shown in the inset of Fig. 2(b). The $\chi(T)$ of LiFe(MoO$_4$)$_2$ can be well fitted to the Curie-Weiss law $\chi = C/(T − \theta_{CW})$, as also shown in Fig. 2(a). Our fitting in the temperature range between 100 K and 300 K yields the Curie constant $C$~4.53 emu K/mol and negative Curie-Weiss temperature $\theta_{CW}$~-52.85 K, suggesting dominant AFM interactions between Fe's spins. The effective moment per $Fe^{3+}$ of 6.02 $\mu_B$ is very close to spin-only moment (5.92 $\mu_B$) for high-spin $Fe^{3+}$ ($S^z$ =5/2, $L$=0).

Figure 2(b) shows the magnetic susceptibility under a high magnetic field (6 T). There is a maximum of $\chi$ at $T_{max}$ ~ 32.8 K, which may due to the ordering of short-range magnetic correlation for this low-dimensional spin system. The peak of $d\chi/dT$ appears near $T_N$~23.8 K, in agreement with aforementioned long-range AFM magnetic ordering (LRO). In addition, Figure 2(c) shows the magnetization ($M$) as a function of magnetic field ($H$) at various temperatures 2 K, 10 K, and 50 K. The $M(H)$ curves below $T_N$ show nonlinear behavior with the applied magnetic field. The largest $M$ value is only ~0.4 $\mu_B$/Fe when magnetic field is up to 6.5 Tesla at 2 K, suggesting the canting AFM state under strong magnetic fields.

We also measured the $T$-dependence of its heat capacity ($C_p$), as shown in the Fig. 3(a). A clear $\lambda$ shaped peak was observed around $T$ ~ 23.8 K, which is one more indication of the long-range magnetic ordering. Considering the fact that LiFe(MoO$_4$)$_2$ is an insulator, the specific heat mainly contains the contributions from both magnons and phonons. At high temperatures, $C_p$ is fully dominated by the phonon excitation contribution. Thus, the $C_p$ of isostructural and non-magnetic LiGa(MoO$_4$)$_2$ [red curve in Fig. 3(a)] was measured and used to subtract the phonon contribution from LiFe(MoO$_4$)$_2$. Then the obtained magnetic heat capacity $C_M$ is shown in Fig. 3(b). The broad hump in both magnetic susceptibility and specific heat around ~ 10 K might originate from the temperature-dependent exchange effects. As suggested in Ref. 33, below $T_N$, the populations and energies of Zeeman levels due to the temperature-dependent exchange field, would be more significant if the spin quantum number is large, as here the spin number $S=5/2$. The magnetic entropy ($S_M$) is estimated through integrating $C_M/T$ over $T$, giving a saturation value of $S_M \sim$ 11.9418 J/mol K at 80 K as shown in Fig. 3(b). This value is very close to the calculation value of total spin entropy of $R\ln(2S+1) = 14.897$ J/mol K ($R = 8.314$ J/mol K). The entropy gain at $T_N$ is only ~48.8% of the total magnetic entropy, indicating the existence of short-range magnetic correlations above $T_N$, which is a characteristic of low dimensional magnets.

To reveal the AFM spin structure of LiFe(MoO$_4$)$_2$, NPD pattern was collected at 1.5 K. As shown in Fig. 4(b), a series of new Bragg peaks indicated by the second row of vertical bars show up compared to the data collected at 40 K. The inset displays a zoomed view of low-$Q$ region to show the peaks more clearly. No obvious change of lattice parameters was observed. In Fig. 4(c), the difference between the 40 and 1.5 K patterns clearly shows the details of these magnetic Bragg peaks. The observed magnetic peaks can be well indexed with a commensurate propagation vector $\mathbf{k} = (0, 0.5, 0)$, distinct from the incommensurate one $\mathbf{k} = (0.890, 0, 0.332)$ seen in LiFe(WO$_4$)$_2$ and a noncollinear triangular spin order seen in RbFe(MoO$_4$)$_2$.

Fig. 4(a) exhibits the temperature evolution of the peak intensity of the (0, -0.5, 1) magnetic Bragg peak. A fit to the power-law $I = A(T_c - T)^{2\beta}$ over the temperature range 8 K to 29 K yields $T_c = 24.1(2)$ K and a critical exponent $\beta = 0.218$. The obtained $\beta$ is close to the expected value for 2D-XY model (~0.23) expected for layered magnetic structures [34].

Representation analysis constrains the possible magnetic structures to the basis vectors associated with an irreducible representation (IR) of the crystal space group and $k = (0, 0.5, 0)$. We used *SARAh* software to perform the analysis on magnetic Fe$^{3+}$ ions. There are two possible irreducible representation allowed for the Fe$^{3+}$ ion at the 2$i$ Wyckoff position, corresponding to $\Gamma_1$ and $\Gamma_2$ in the Kovalev numbering scheme, as listed in Table II. Three basis vectors are allowed for each representation. While the $\Gamma_1$ is ruled out since it cannot reproduce the correct magnetic intensities, the $\Gamma_2$ spin model gives us good fitting results as shown in Fig. 4(b). The corresponding magnetic structure is presented in Fig. 5 with the refined amplitude of the magnetic moment is

4.23(1) $\mu_B$. The projections of the moment on the crystallographic axes are: ($m_a$, $m_b$, $m_c$) = (0.582, 4.2, 1.296) $\mu_B$. This value is only slightly smaller than the expected value for the spin S = 5/2, but it is still reasonable considering the low dimensionality magnetic interactions.

To further understand the magnetic properties, DFT calculations have been performed. Various possible magnetic arrangements were checked, such as, ferromagnetic (FM), AF1-AFM, AF2-AFM, and AF3-AFM, as shown in Fig. 6. As shown in Fig. 7, the AF2-AFM configuration has the lowest energy despite the choice of $U_{eff}$, in agreement with our neutron experimental result. The calculated local magnetic moment of AF2-AFM state is 4.22 $\mu_B$/Fe at $U_{eff}$=4 eV, which is quite closed to our neutron experiments (4.23 $\mu_B$/Fe). In fact, $U_{eff}$=4 eV is a proper choice to describe Fe's 3$d$ orbitals according to previous studies [17, 35], which will be adopted as the default one. The optimized lattice constants of AF2-AFM state are $a$=6.748 Å, $b$=7.233 Å and $c$= 7.113 Å, which are quite close to our neutron experiments ($a$=6.751 Å, $b$=7.203 Å, $c$=7.1702 Å at 1.5 K).

Furtheremore, the magnetism of LiFe(Mo$_4$)$_2$ can be described using a Heisenberg model:

$$H = -J_1 \sum_{<ij>} S_i \cdot S_j - J_2 \sum_{[kl]} S_k \cdot S_l - J_3 \sum_{\{mn\}} S_m \cdot S_n, \quad (1)$$

where $J_1$, $J_2$ and $J_3$ are the exchange couplings [as indicated in Fig.1(b)] between iron spins $S$'s. Using the optimized ground state configuration, the exchange coefficients are extracted from DFT calculations: $J_1$=-18.03 meV, $J_2$=2.06 meV and $J_3$=2.07 meV, respectively. The strongest negative exchange $J_1$ indicates the AFM coupling between the nearest neighbor iron spins. Those longer distance exchanges $J_2$ and $J_3$ are much weaker: the positive $J_2$ implies FM interaction along $b$-axis, and the positive $J_3$ means that the magnetic coupling between near iron is FM exchange. In short, our theoretical calculation confirms the magnetic ground state of LiFe(MoO$_4$)$_2$.

The atomic-projected density of states (DOS) of AF2-AFM state is shown in Fig. 8. It is clear that this system is an insulator with an indirect gap of 2.46 eV. According to DOS, the topmost valence bands are mainly contributed by the hybrid O's 3$p$ orbitals and Fe's *3d* orbitals (lower Hubbard bands), while the lowest conduction bands are mainly from Fe's *3d* orbitals (i.e. upper Hubbard bands). The slight reduction of local magnetic moment from the ideal 5 $\mu_B$/Fe is due to such hybridization.

## IV. Conclusion

In summary, the physical properties of the spin-5/2 double molybdate LiFe(MoO$_4$)$_2$ have been systematically investigated experimentally and theoretically. Our magnetic susceptibility and heat capacity measurements found an antiferromagnetic long range ordering at $T_N$ ~ 23.8 K, which was further confirmed by neutron diffraction. Its antiferromagnetic magnetic structure with the commensurate propagation vector ***k*** = (0, 0.5, 0) was also been determined. Our DFT calculations further verified the magnetic ground state. Short range magnetic correlation was also evidenced above the Néel temperature.

**Acknowledgement**

This work was supported by the National Key Research Projects of China (Grant No. 2016YFA0300101), and the National Natural Science Foundation of China (Grant Nos. 11704109, 11834002). The research at Oak Ridge National Laboratory's High Flux Isotope Reactor was sponsored by the Scientific User Facilities Division, Office of Basic Energy Sciences, US Department of Energy. Most calculations were supported by National Supercomputer Center in Guangzhou (Tianhe II).

**Figure Captions:**

Fig. 1. (a) Projection in the bc plane of the crystal structure of LiFe(MoO$_4$)$_2$. Red: Fe; blue: W; black: Li; gray: O. (b) The framework of Fe ions and the magnetic exchange paths $J_1$/$J_2$/$J_3$. (c) The XRD pattern measured at room T and the corresponding Rietveld fit. (d) Neutron powder diffraction pattern collected at 40 K with a neutron wavelength of λ = 2.41 Å.

Fig. 2. (a) Temperature dependence of magnetic susceptibility (left y axis) and its inverse (right y axis) of LiFe(MoO$_4$)$_2$ measured under 0.1 T field. (b) The temperature evolution of magnetic susceptibility under various magnetic fields. The inset shows the amplified view of the derivatives of χ around the phase transition temperature. (c) The moment size of Fe$^{3+}$ ion as a function of magnetic field at various temperatures.

Fig. 3. (a) Heat capacity of magnetic LiFe(MoO$_4$)$_2$ and nonmagnetic LiGa(MoO$_4$)$_2$ measured under 0 T magnetic field. (b) The magnetic contribution of the specific heat $C_{mag}$ as a function of temperature acquired by subtracting the phonon contribution using nonmagnetic LiGa(MoO$_4$)$_2$ (left y axis) and the magnetic entropy $S_{mag}$ as a function of temperature (right y axis).

Fig. 4. (a) The temperature evolution of the (0, -0.5, 1) magnetic Bragg peak. (b) neutron powder diffraction pattern collected at 1.5 K with a neutron wavelength of λ = 2.41 Å. The inset shows the enlarged area of the low Q region to clearly exhibit the magnetic Bragg peaks. (c) Comparing of neutron powder diffraction patterns collected at 1.5 and 40 K.

Fig. 5. Magnetic structure of LiFe(MoO$_4$)$_2$ determined from the neutron diffraction patterns.

Fig. 6. Sketch of possible spin configurations in LiFe(MoO$_4$)$_2$ lattice.

Fig. 7. DFT results of single LiFe(MoO$_4$)$_2$ as a function of $U_{\text{eff}}$. (a) Energy (per Fe) of various magnetic orders. The FM state is taken as the reference. (b) Local magnetic moment of Fe calculated within the default Wigner-Seitz sphere. (c) Band gaps. (d) Relaxed lattice constants of AF2 states.

Fig. 8. The atomic-projected density of states (DOS) of AF2 ($U_{\text{eff}}$= 4 eV). Below the Fermi level, the oxygen and irons have large overlap ranges, implying orbital hybridization.

Table I. Refined structural parameters of LiFe(MoO$_4$)$_2$ from neutron powder diffraction data collected at 40 K.

| Atom (Wyck.) | x | y | z | B |
| --- | --- | --- | --- | --- |
| Mo1 (2$i$) | 0.3326(4) | 0.5768(2) | 0.2907(3) | 0.071(3) |
| Mo2 (2$i$) | 0.8221(6) | -0.0395(5) | 0.2272(5) | 0.071(3) |
| Fe (2$i$) | 0.4001(4) | 0.0996(4) | 0.31752(4) | 0.073(4) |
| Li (2$i$) | 0.7580(2) | 0.4418(2) | 0.2557(2) | 0.803(2) |
| O1 (2$i$) | 0.4166(7) | 0.8407(73) | 0.3907(6) | 0.342(2) |
| O2 (2$i$) | 0.2563(5) | 0.5731(65) | 0.03978(6) | 0.342(2) |
| O3 (2$i$) | 0.0891 (7) | 0.4879(60) | 0.3351(6) | 0.342(2) |
| O4 (2$i$) | 0.4795(6) | 0.3805(63) | 0.3539(6) | 0.342(2) |
| O5 (2$i$) | 0.6896(7) | 0.1301(63) | 0.2682(7) | 0.342(2) |
| O6 (2$i$) | 0.7755(6) | -0.0499(70) | -0.0271(6) | 0.342(2) |
| O7 (2$i$) | 0.7082(6) | -0.2787 (60) | 0.2690(9) | 0.342(2) |
| O8 (2$i$) | 1.1138(6) | 0.0417 (63) | 0.35852(6) | 0.342(2) |

SP: $P\text{-}1$, $a = 6.751\,(1)$ Å, $b = 7.203\,(2)$ Å, $c = 7.1702(1)$ Å, $\alpha = 90.6739\,(17)°$, $\beta = 110.2274(14)°$, $\gamma = 105.5793(17)°$,

$R_p = 3.00\%$, $R_{wp} = 5.61\%$

Table II. Basis vectors for the space group P-1 with the propagation vector k = (0, 0.5, 0). The decomposition of the magnetic representation for the Fe site (0.4027, 0.102, 0.318) is $\Gamma_{Fe} = 3\Gamma_1 + 3\Gamma_2$. The atoms of the nonprimitive basis are defined according to 1: (0.4027, 0.102, 0.318), 2:(0.597, 0.897, 0.682).

| IR | BV | Fe1 (0.4027, 0.102, 0.318) | Fe2 (0.597, 0.897, 0.682) |
|---|---|---|---|
| $\Gamma_1$ | $\psi_1$ | (1, 0, 0) | (-1, 0, 0) |
| | $\psi_2$ | (0, 1, 0) | (0, -1, 0) |
| | $\psi_3$ | (0, 0, 1) | (0, 0, -1) |
| $\Gamma_2$ | $\psi_4$ | (1, 0, 0) | (1, 0, 0) |
| | $\psi_5$ | (0, 1, 0) | (0, 1, 0) |
| | $\psi_6$ | (0, 0, 1) | (0, 0, 1) |

Figure 1

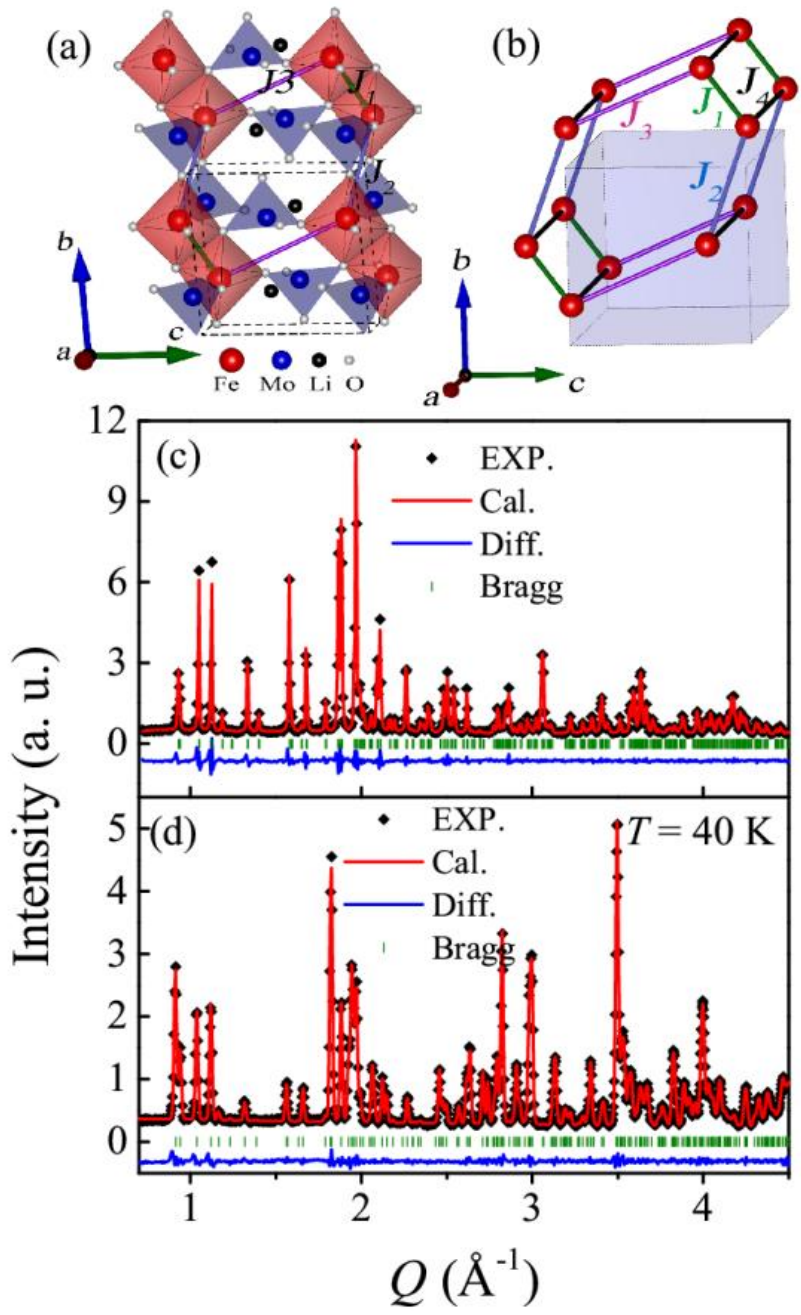

Figure 2

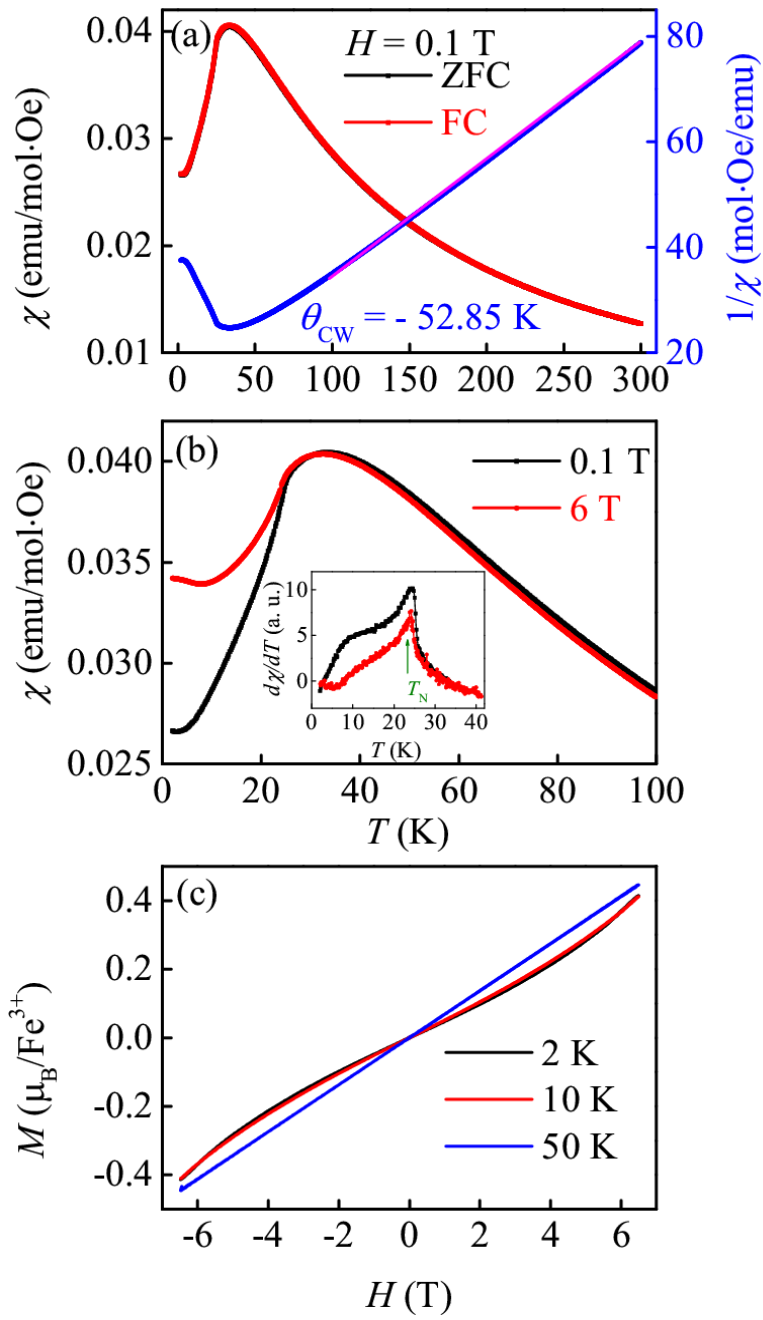

Figure 3

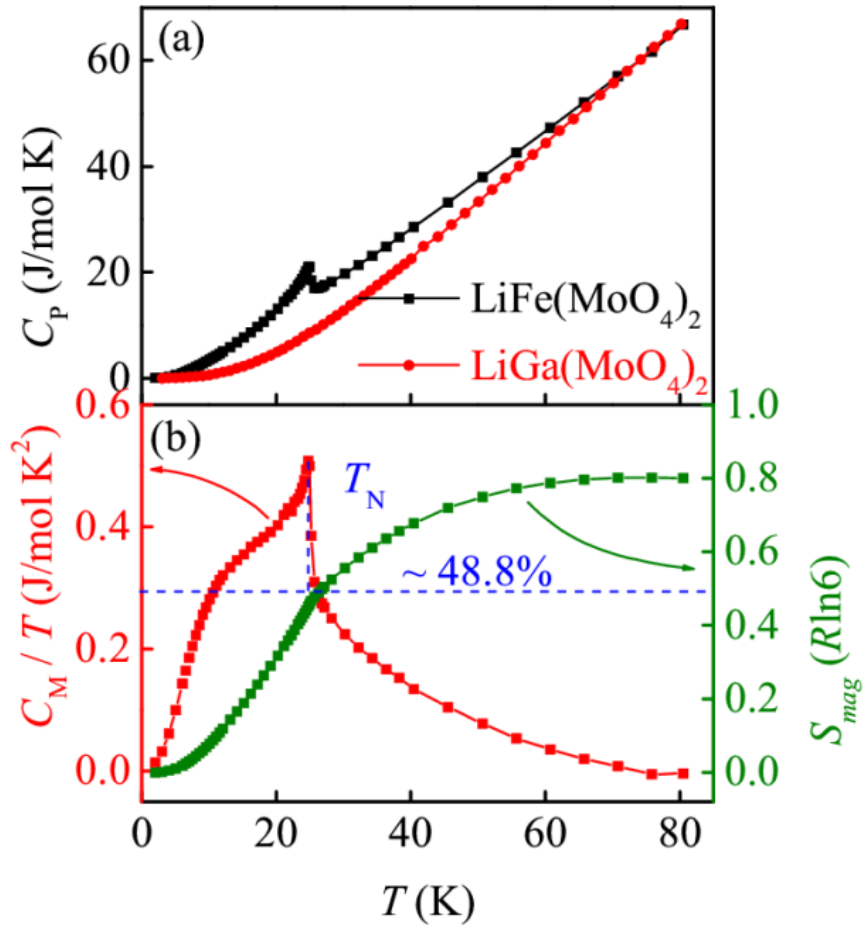

Figure 4

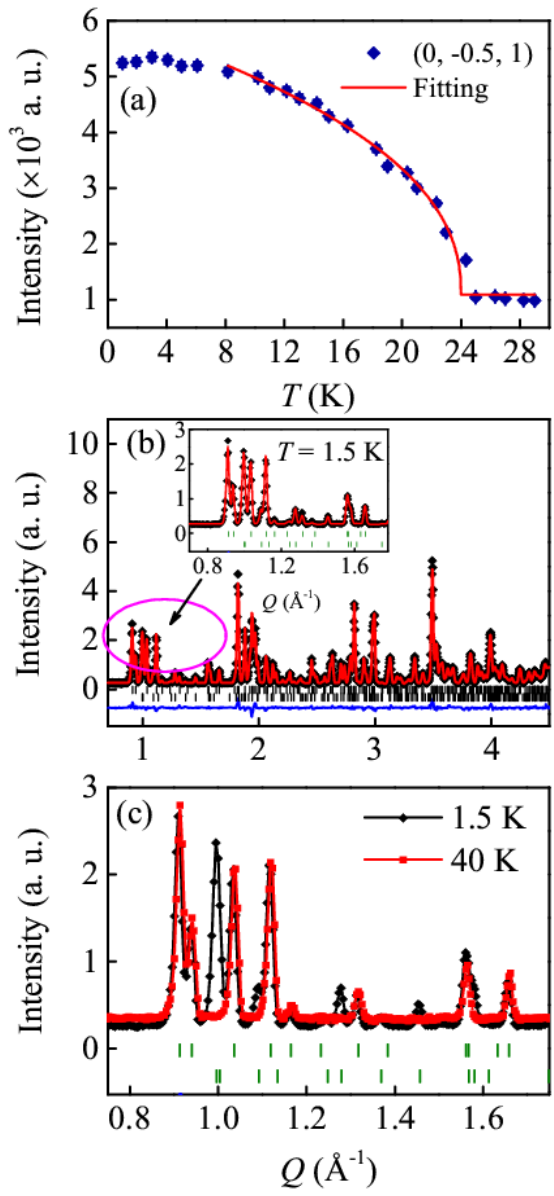

Figure 5

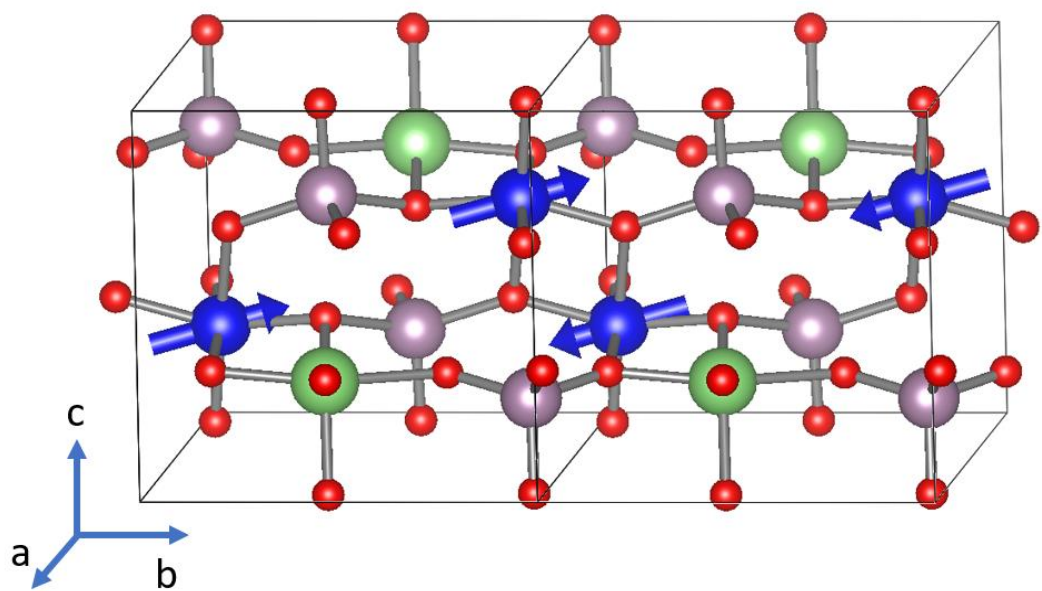

Figure 6

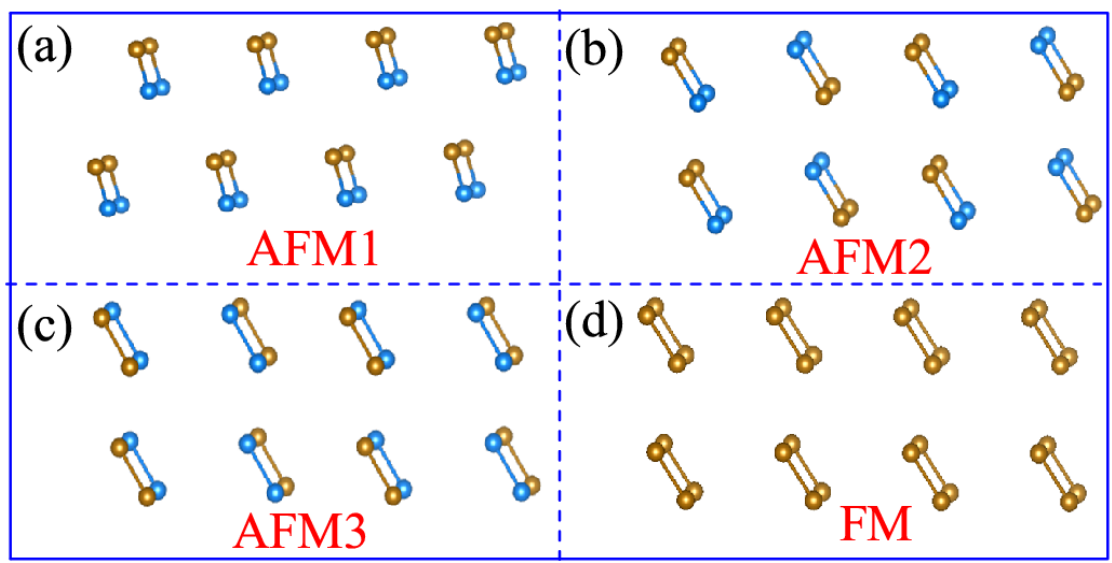

Figure 7

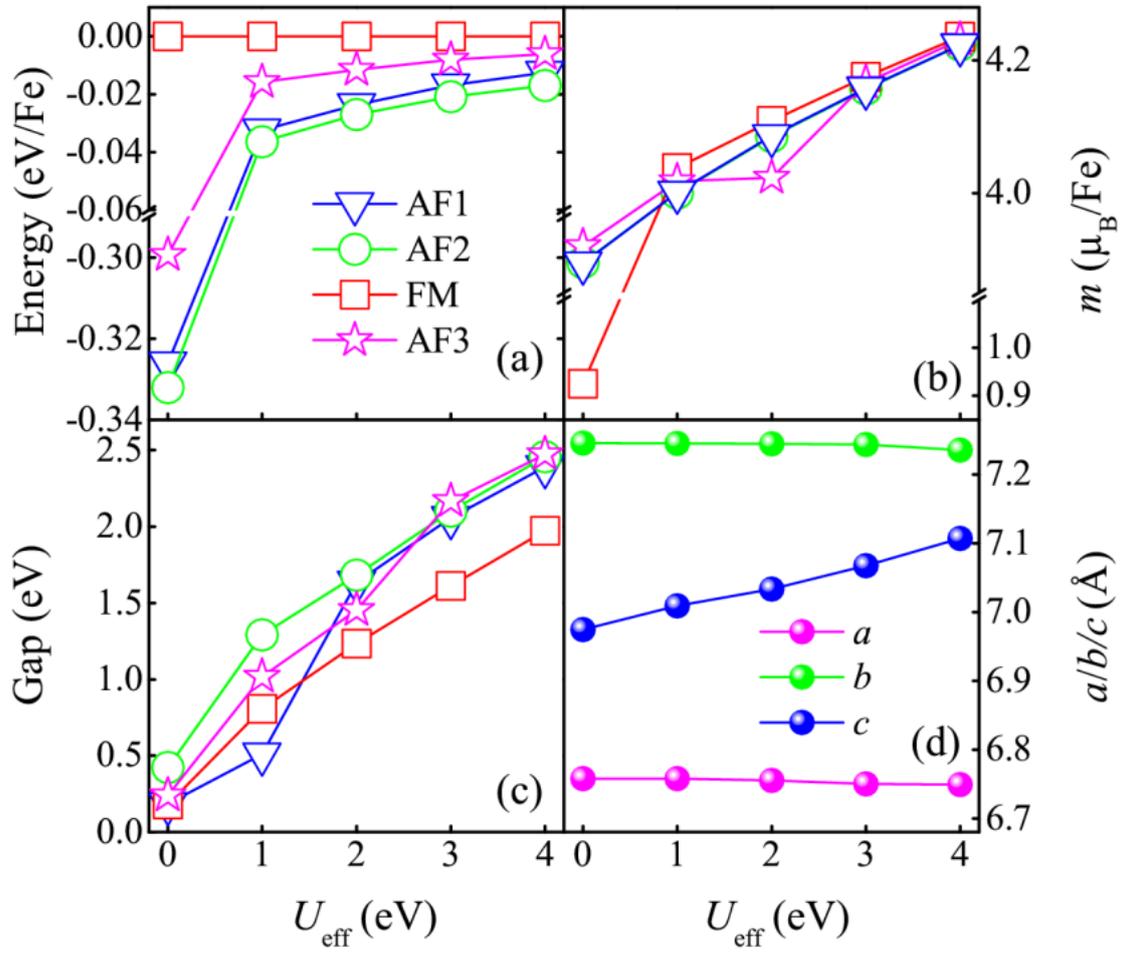

Figure 8

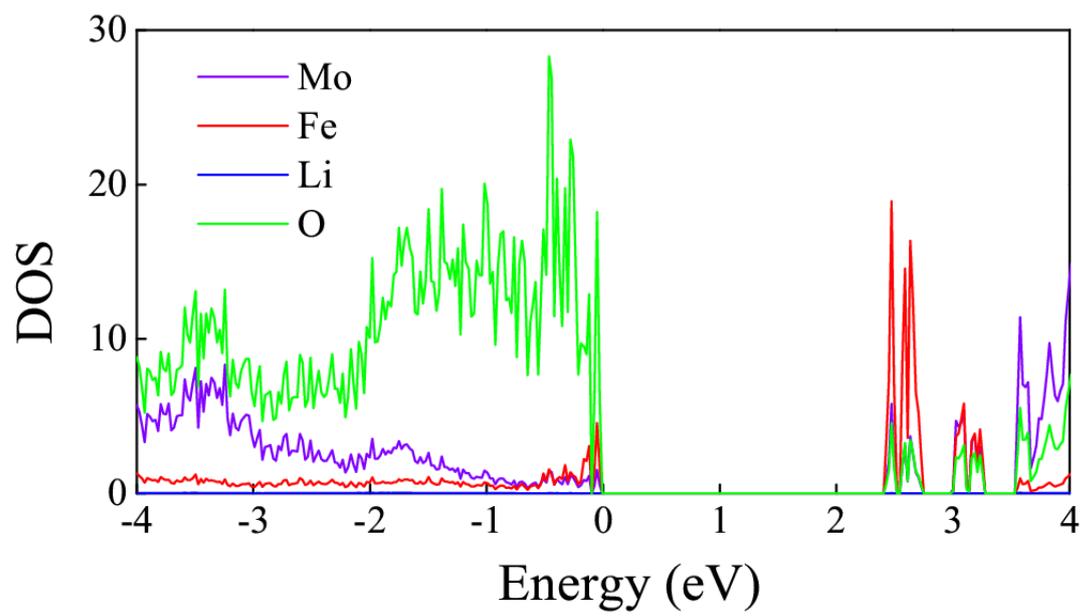